\documentclass{main}
\usepackage{textcomp}
\usepackage{fancyhdr}
\usepackage{lastpage}
\usepackage{pifont}
\usepackage{xcolor}
\usepackage{amsmath}
\usepackage[export]{adjustbox}

\usepackage[numbers,compress]{natbib}
\usepackage{cleveref}
\crefname{lstlisting}{lst.}{listing}
\Crefname{lstlisting}{Lst.}{Listings}

\usepackage{listings}			% package for listings
\newcommand{\red}[1]{\textcolor[rgb]{1,0,0}{#1}}
\newcommand{\forestgreen}[1]{\textcolor[rgb]{0.25,0.53,0.22}{#1}}

\def\gaga{\gamma\gamma}
\def\mumu{\mu^+\mu^-}
\def\tata{\tau^+\tau^-}
\def\ttbar{t\bar{t}}

\def\madgraph{{\sc MadGraph5\_aMC@NLO}}
\def\mgshort{{\sc MG5\_aMC}}
\def\gammaUPC{{\tt gamma-UPC}}
\def\upcgen{{\tt Upcgen}}
\def\cepgen{{\tt CepGen}}
\def\fpmc{{\tt FPMC}}
\def\starlight{{\sc STARlight}}
\def\superchic{{\sc SuperChic}}

\def\pythia{{\tt Pythia}}
\def\herwig{{\tt Herwig}}
\def\be{\begin{equation}}
\def\ee{\end{equation}}
\def\bea{\begin{equation}\begin{aligned}}
\def\eea{\end{aligned}\end{equation}}
\def\cmark{\text{\forestgreen{\ding{51}}}}
\def\xmark{\text{\red{\ding{55}}}}
\newcommand\prompt{{\tt MG5\_aMC>}}

\newcommand{\Photo}{}

\fancypagestyle{firstpagefooter}{%
  \fancyfoot[L]{  \textcopyright \footnotesize This work is an open access article under a Creative Commons Attribution 4.0 International License (https://creativecommons.org/licenses/by/4.0/).}
  \fancyfoot[C]{}  % **Red: This removes the page number from the first page footer**
   
}

\begin{document}

\thispagestyle{firstpagefooter}
\title{\Large Progress in NLO Calculations for gamma-gamma Physics}

\author{\underline{Hua-Sheng Shao}\footnote{Speaker, email: huasheng.shao@lpthe.jussieu.fr} and Lukas Simon\footnote{email: lsimon@lpthe.jussieu.fr}}

\address{
Laboratoire de Physique Th\'eorique et Hautes Energies (LPTHE), UMR 7589, Sorbonne Universit\'e et CNRS, 4 place Jussieu, 75252 Paris Cedex 05, France
}

\maketitle\abstracts{
With the advent of precision measurements of photon-fusion processes in ultraperipheral collisions (UPCs) at facilities, such as RHIC and the LHC, the inclusion of higher-order corrections has become essential. While automated frameworks already make NLO corrections feasible for parton-parton scattering processes, no comparable tools had previously been available for UPC processes in two-photon collisions. In these proceedings, we review some recent explicit NLO computations and present the updated \gammaUPC+\madgraph\ framework, the first automated software that enables NLO-accurate predictions for photon–photon processes in UPCs.
}

%\footnotesize DOI: \url{https://doi.org/xx.yyyyy/nnnnnnnn}


%%%%%%%%%%%%%%%%%%%%%%%%
%% INTRODUCTION       %%
%%%%%%%%%%%%%%%%%%%%%%%%
\section{Introduction}

The study of ultraperipheral collisions (UPCs) at hadron colliders offers the opportunity for probing the structure of the colliding nuclei and can serve as a precision test of quantum chromodynamics (QCD). UPCs grant access to a wide range of processes within the Standard Model (SM) and Beyond-the-Standard Model (BSM)~\cite{Shao:2022cly}. A highlight is the recent observation of $\tau$-pair production in UPCs at the CMS experiment, providing the best constraint on the anomalous magnetic moment of the $\tau$ lepton, $a_\tau=0.0009(^{+0.0016}_{-0.0015})_\mathrm{syst.}(^{+0.0028}_{-0.0027})_\mathrm{stat.}$~\cite{CMS:2024qjo}. This was measured in the scattering of two photons in proton-proton collisions. Given that the coherent photon flux from the electromagnetic field surrounding a nucleus scales as the square of its electric charge $Z^2$, two-photon interactions in nucleus-nucleus collisions benefit from an enhanced luminosity that grows as $Z^4$, making it particular interesting to study these processes in heavy-ion collisions.\\
Modern hadron colliders, such as the Relativistic Heavy Ion Collider (RHIC) at BNL and the Large Hadron Collider (LHC) at CERN, provide an ideal environment for investigating these photon-induced processes. Over the past two decades, these facilities have driven major experimental and theoretical progress. With growing community interest, advancing detector technologies, and improving analysis methods, UPCs are poised to deliver deeper insights into the fundamental principles of particle physics in the near future. To keep track with the experimental developments, theorists are required to refine and extend their predictions. Several public frameworks provide tools to model two-photon luminosities, including \starlight~\cite{Klein:2016yzr}, \superchic~\cite{Harland-Lang:2020veo}, \fpmc~\cite{Boonekamp:2011ky}, \cepgen~\cite{Forthomme:2018ecc}, \upcgen~\cite{Burmasov:2021phy}, and \gammaUPC~\cite{Shao:2022cly}. Despite this progress, most predictions are still limited to leading-order (LO) precision. Extending to next-to-leading order (NLO) accuracy requires modifications to the infrared-divergence subtraction methods implemented in the underlying event generators. These methods are typically tailored for inclusive reactions in hadronic collisions, rather than for initial states with coherent photons. As a result, NLO corrections, essential for reliable comparisons with experimental data, are computed individually for each process in customized settings. Currently, only a few processes have been studied beyond LO. These include $\gaga \to \ttbar$ at NLO QCD~\cite{Shao:2022cly}, $\gaga \to \mumu$ and $\gaga \to \tata$ with NLO quantum electrodynamics (QED) corrections for bare final-state leptons~\cite{Shao:2024dmk}, and NLO electroweak (EW) corrections for $\gaga \to \tata$ with bare leptons~\cite{Jiang:2024dhf}. In addition, EW corrections for decayed $\tau$-pair production $\gaga \to \tata \to e^+\mu^-\bar{\nu}_\tau\nu_\tau\bar{\nu}_\mu\nu_e$ have recently been computed in the narrow-width approximation (NWA), including spin correlations~\cite{Dittmaier:2025ikh}. Furthermore, NLO QCD and QED corrections for the loop-induced light-by-light (LbL) scattering process $\gaga \to \gaga$ have been calculated in refs.~\cite{AH:2023ewe,AH:2023kor}.\\
With the recent upgrade of the \madgraph\ framework~\cite{Alwall:2014hca,Frederix:2018nkq} (henceforth dubbed \mgshort) in conjunction with the \gammaUPC\ code, NLO computations of photon-photon scattering processes have, for the first time, been automated for a wide class of UPC SM processes~\cite{Shao:2025bma}. This includes both NLO QCD and NLO EW corrections.

%%%%%%%%%%%%%%%%%%%%%%%%
%% SECTION 2          %%
%%%%%%%%%%%%%%%%%%%%%%%%
\section{Impact of NLO Corrections}

In the framework of collinear factorization, the cross section of a photon–photon scattering process in UPCs of two nuclei, $\mathrm{A}_1$ and $\mathrm{A}_2$, producing a final state $X$, is given by:

\be\label{eq:two-photon}
\sigma(\mathrm{A}_1\; \mathrm{A}_2\,\xrightarrow{\gaga} \mathrm{A}_1 \; X \; \mathrm{A}_2)=
\int \frac{\mathrm{d}E_{\gamma_1}}{E_{\gamma_1}} \frac{\mathrm{d}E_{\gamma_2}}{E_{\gamma_2}} \, \frac{\mathrm{d}^2N^{(\mathrm{A}_1\mathrm{A}_2)}_{\gamma_1/Z_1,\gamma_2/Z_2}}{\mathrm{d}E_{\gamma_1}\mathrm{d}E_{\gamma_2}} \sigma_{\gaga\to X}(W_{\gaga})
\ee
\noindent
This expression is a convolution of two distinct components.

The first one is the effective two-photon luminosity
\be\label{eq:2photonintegral}
\frac{\mathrm{d}^2N^{(\mathrm{A}_1\mathrm{A}_2)}_{\gamma_1/Z_1,\gamma_2/Z_2}}{\mathrm{d}E_{\gamma_1}\mathrm{d}E_{\gamma_2}} =  \int{\mathrm{d}^2\pmb{b}_1\mathrm{d}^2\pmb{b}_2\,P_\mathrm{no\ inel}(\pmb{b}_1,\pmb{b}_2)\,N_{\gamma_1/Z_1}(E_{\gamma_1},\pmb{b}_1)N_{\gamma_2/Z_2}(E_{\gamma_2},\pmb{b}_2)}\,,
\ee
which gives the probability to find two colliding photons $\gamma_1$ and $\gamma_2$ with energies $E_{\gamma_1}$ and $E_{\gamma_2}$ emitted from nuclei $\mathrm{A}_1$ and $\mathrm{A}_2$ of charges $Z_1$ and $Z_2$. This is in full analogy to the parton luminosity in hard scattering processes, which is defined as the convolution of two parton distribution functions (PDFs). The evaluation of this two-photon luminosity requires both the survival probability of the colliding nuclei, encoded in the no-inelastic-interaction probability density $P_\mathrm{no\ inel}$, and the photon number density $N_{\gamma/Z}$ as functions of photon energy $E_\gamma$ and impact parameter $\pmb{b}$. These must be integrated over the transverse distances $\pmb{b}_1$ and $\pmb{b}_2$, which describe the points relative to the centres of $\mathrm{A}_1$ and $\mathrm{A}_2$ at which the photons collide. We refer the works in refs.~~\cite{Shao:2022cly,Klein:2016yzr,Harland-Lang:2020veo,Boonekamp:2011ky,Forthomme:2018ecc,Burmasov:2021phy} and references therein for modelling this effective two-photon luminosity.

\begin{figure}
    \centering
    \includegraphics[width=0.7\linewidth]{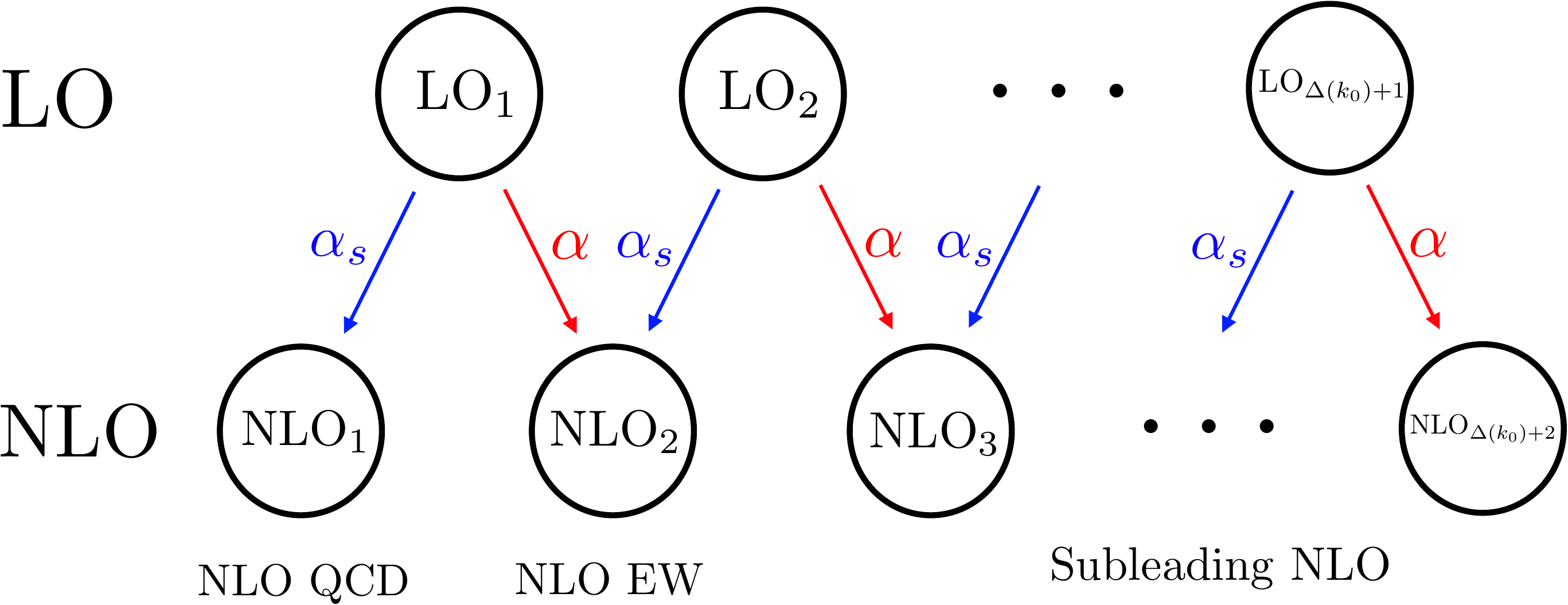}
    \caption{A graphical illustration of QCD (blue, right-to-left arrows) and EW (red, left-to-right arrows) corrections to a generic process. Each circle represents a specific contribution at a fixed order in terms of the QCD and EW coupling constants, while the black ellipses indicate the potential presence of additional cross section contributions. Figure taken from ref.~\protect\cite{Shao:2025fwd}.}
    \label{fig:cross_section_decomposition}
\end{figure}

The second piece in \cref{eq:two-photon} is the partonic hard scattering cross section, $\mathrm{d}\sigma_{\gamma\gamma\to X}$, depending on the photon–photon centre-of-mass energy $W_{\gaga}=2\sqrt{E_{\gamma_1}E_{\gamma_2}}$. Since the incoming coherent photons have almost zero virtuality, these quasi-real photons can be described by the Equivalent Photon Approximation (EPA)~\cite{Fermi:1924tc,vonWeizsacker:1934nji,Williams:1934ad}. Due to the perturbative nature of the partonic cross section, this term can be expanded in the strong and electroweak coupling constants, $\alpha_s$ and $\alpha$. This results in the expression
\bea\label{eq:Sigmaexp0}
\sigma_{\gaga\to X}(W_{\gaga})&=\alpha_s^{c_s(k_0)}\alpha^{c(k_0)}\sum_{p=0}^{+\infty}{\sum_{q=0}^{\Delta(k_0)+p}{\Sigma_{k_0+p,q}\alpha_s^{\Delta(k_0)+p-q}\alpha^q}}\\
&=\sigma^{({\rm LO})}(W_{\gaga})+\sigma^{({\rm NLO})}(W_{\gaga})+\ldots\,,
\eea
where the LO term includes all contributions with the minimal total power of couplings, corresponding to $p=0$ in the above equation. The LO term can be further decomposed following the hierarchy $\alpha_s>\alpha$,
\bea\label{eq:decomposition_lo}
\sigma^{({\rm LO})}(W_{\gaga})&=\alpha_s^{c_s(k_0)}\alpha^{c(k_0)}\sum_{q=0}^{\Delta(k_0)}{\Sigma_{k_0,q}\alpha_s^{\Delta(k_0)-q}\alpha^q}\\
&=\Sigma_{\mathrm{LO}_1}+\ldots+\Sigma_{\mathrm{LO}_{\Delta(k_0)+1}}\,,\\
\eea
with $\Sigma_{\mathrm{LO}_1}$ proportional to the highest power of $\alpha_s$ and the lowest power of $\alpha$, while successive terms iteratively decreasing in the power of $\alpha_s$ and increasing in the power of $\alpha$. This ordering naturally imposes $\Sigma_{\mathrm{LO}_1}>\Sigma_{\mathrm{LO}_2}>\dots>\Sigma_{\mathrm{LO}_{\Delta(k_0)+1}}$.\footnote{This inequality relies on the na\"ive assumption that the couplings alone determine the relative importance of the contributions, whereas in practice formally smaller terms can sometimes dominate due to kinematic or dynamical effects~\cite{Shao:2025fwd}.} In a similar fashion, the NLO terms can be written as
\bea\label{eq:decomposition_nlo}
\sigma^{({\rm NLO})}(W_{\gaga})&=\alpha_s^{c_s(k_0)}\alpha^{c(k_0)}\sum_{q=0}^{\Delta(k_0)+1}{\Sigma_{k_0+1,q}\alpha_s^{\Delta(k_0)+1-q}\alpha^q}\\
&=\Sigma_{\mathrm{NLO}_1}+\ldots+\Sigma_{\mathrm{NLO}_{\Delta(k_0)+2}}\,,
\eea
where each $\Sigma_{\mathrm{NLO}_i}$ contains one additional power of $\alpha_s$ compared to the corresponding LO term $\Sigma_{\mathrm{LO}_i}$. This means that $\Sigma_{\mathrm{NLO}_i}$ ($\Sigma_{\mathrm{NLO}_{i+1}}$) describes the NLO QCD (EW) corrections to the term $\Sigma_{\mathrm{LO}_i}$. A schematic illustration of this decomposition is shown in \cref{fig:cross_section_decomposition}. Conventionally, the term $\Sigma_{\mathrm{LO}_1}$ is identified as the LO contribution, while $\Sigma_{\mathrm{NLO}_1}$ and $\Sigma_{\mathrm{NLO}_2}$ are referred to as NLO QCD and NLO EW corrections, respectively. For these proceedings we stick to these conventions, if not stated otherwise.

\subsection{$\gaga \to \mumu$}\label{subsec:mumu}

\begin{figure}
    \includegraphics[width=0.33\textwidth]{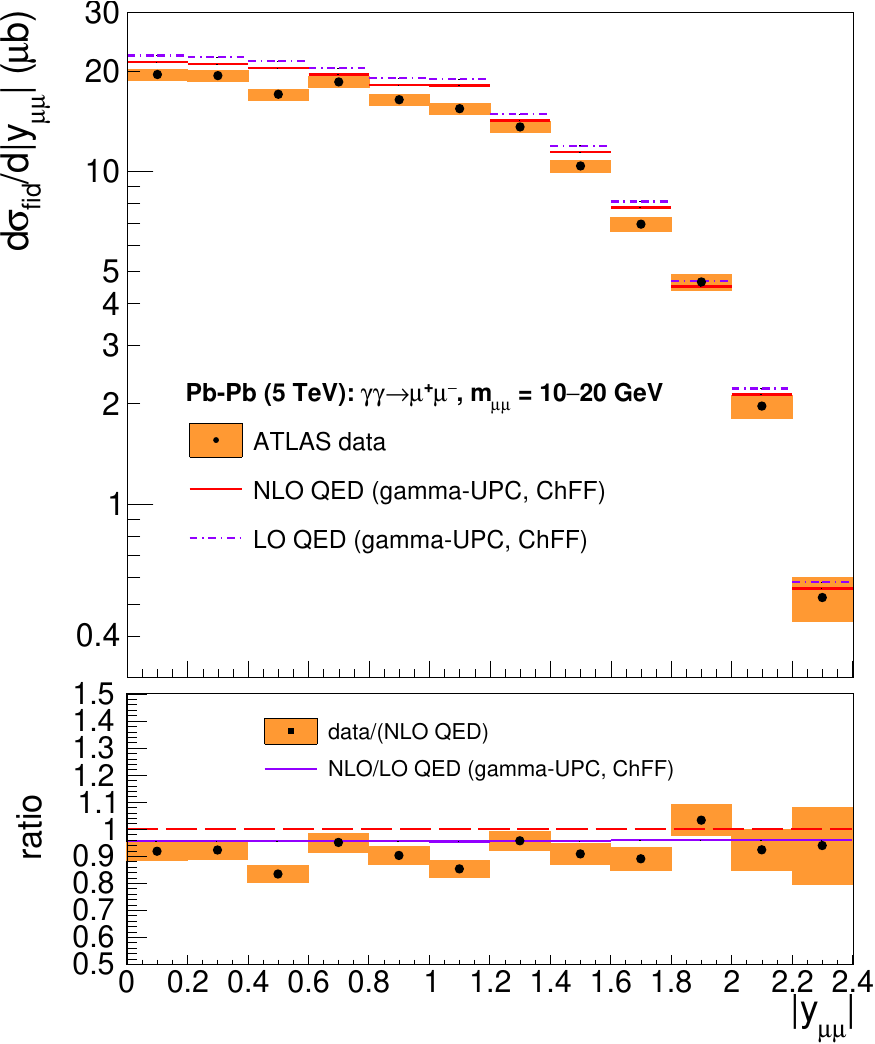}
    \includegraphics[width=0.33\textwidth]{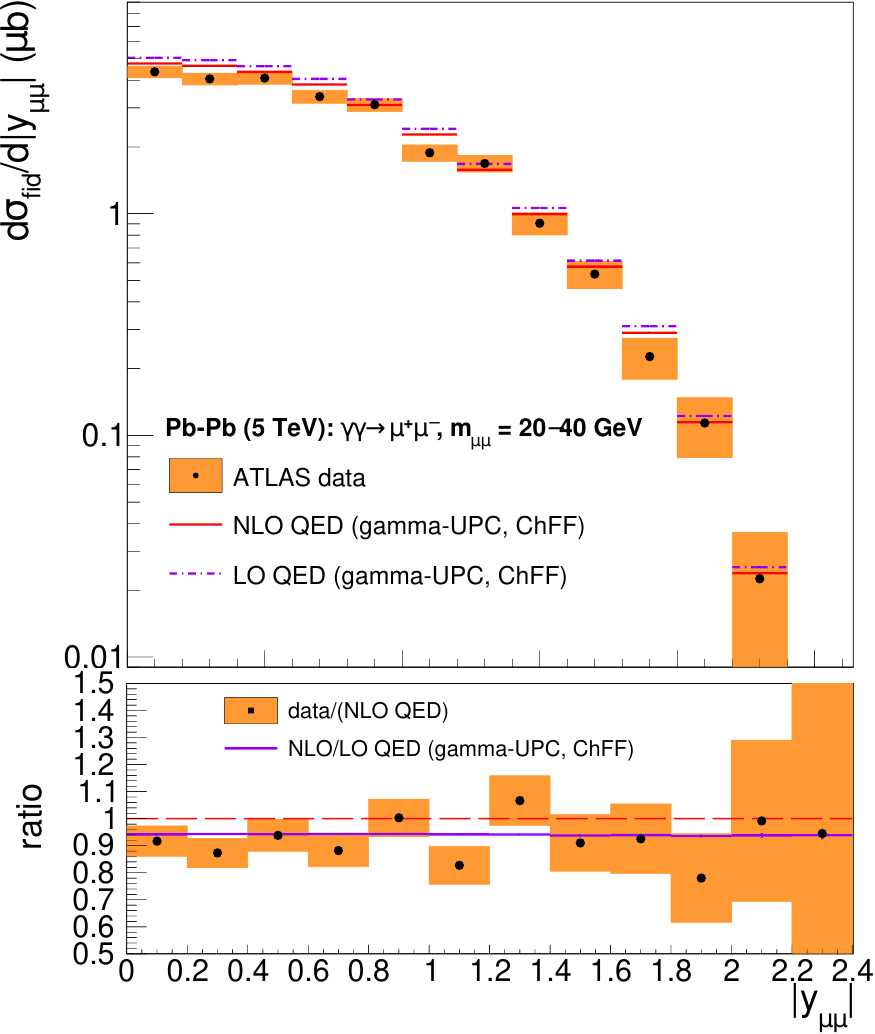}
    \includegraphics[width=0.33\textwidth]{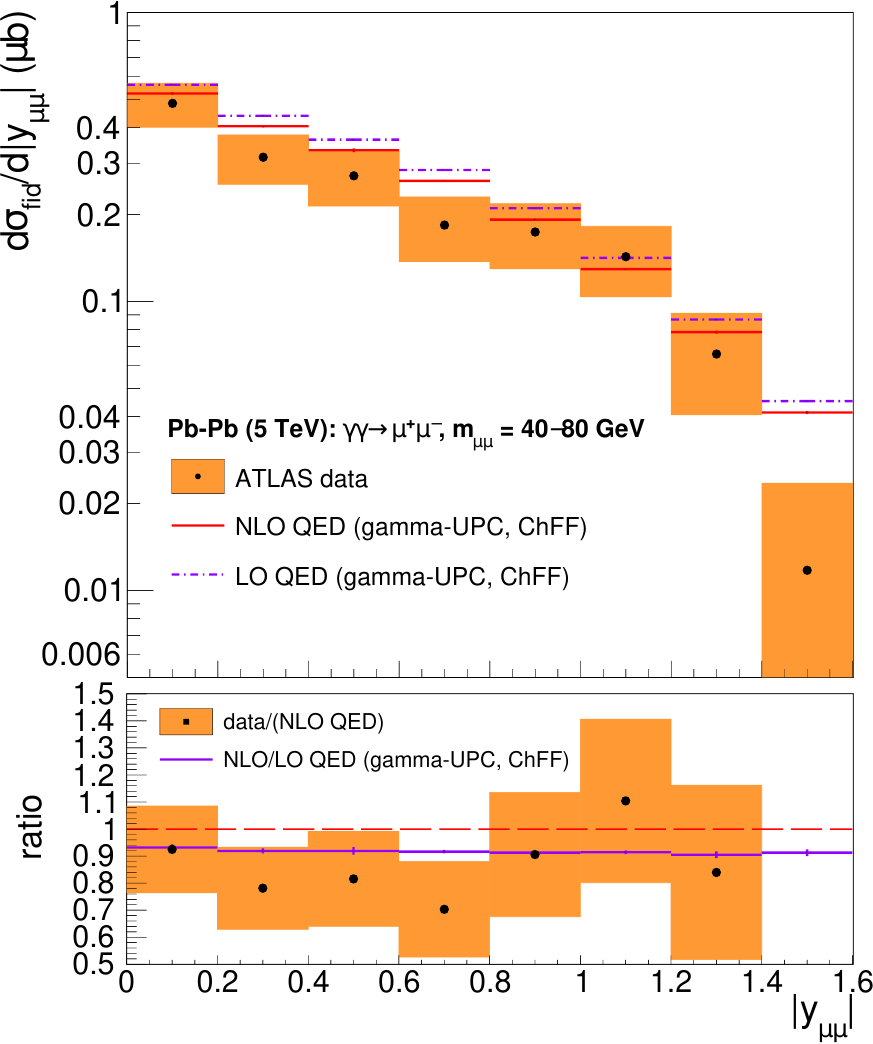}
    \caption{Differential cross section as a function of dimuon absolute rapidity, $\mathrm{d}\sigma/\mathrm{d}|y_{\mumu} |$, for exclusive dimuon production in Pb-Pb
UPCs at $\sqrt{s_{NN}} = 5.02\,\mathrm{TeV}$ in different ranges of invariant mass. The data (black dots with orange uncertainties)~\protect\cite{ATLAS:2020epq} are compared with LO (blue dashed histogram) and NLO (red histogram) QED \gammaUPC\ predictions with ChFF photon fluxes. The lower panels of each plot show the corresponding data/NLO ratios (black dots with orange uncertainties) and NLO/LO $K$ factors (violet histogram). Figure taken from ref.~\protect\cite{Shao:2024dmk}.}
    \label{fig:aa2mumu}
\end{figure}

The importance of NLO corrections for UPC processes becomes evident even in a simple case such as muon-pair production, $\gaga \to \mumu$. Muons can be reconstructed very precisely with a unique experimental signal, allowing them to be directly detected as bare particles in the muon chambers. Consequently, theoretical predictions for bare-muon production cross sections must retain the non-zero muon mass in the calculations. This was explicitly studied in ref.~\cite{Shao:2024dmk}, where predictions were obtained for p–p and Pb–Pb collisions at various LHC energies using the \gammaUPC\ framework interfaced with \mgshort\ for event generation. The study includes NLO QED corrections and employs two different models for the photon fluxes, namely the electric-dipole (EDFF) and charge (ChFF) form factors. A comparison with data shows that the ChFF model reproduces the measurements more accurately, and is therefore preferred. Moreover, the inclusion of NLO QED corrections significantly improves the theory-data comparison. As shown in \cref{fig:aa2mumu}, the absolute rapidity spectra of the dimuon system in Pb–Pb UPCs at $\sqrt{s_{NN}}=5.02\,\mathrm{TeV}$ were computed in various fiducial phase-space regions and compared with ATLAS measurements~\cite{ATLAS:2020epq}. While LO predictions fall short of describing the data, the NLO QED results exhibit better agreement across all kinematic ranges within experimental uncertainties, highlighting the necessity of higher-order corrections.

\subsection{$\gaga \to \tata$}\label{subsec:tata}

The process $\gaga \to \tata$ is of particular importance as it provides a direct probe of the $\gamma\tau\tau$ vertex and thereby enables a determination of the anomalous magnetic moment of the $\tau$ lepton. The best current constraint on this quantity was obtained by the CMS experiment in $\sqrt{s}=13\,\mathrm{TeV}$ p-p UPCs, as mentioned earlier. A precise knowledge of the NLO EW corrections is essential for a robust interpretation of this measurement. The QED corrections have been computed in ref.~\cite{Shao:2024dmk}, while the full EW corrections were presented in ref.~\cite{Jiang:2024dhf}. An interesting difference between the two studies lies in their choice of the EW renormalisation scheme. In the former article, the $\alpha(0)$ renormalisation scheme was used, where the fine structure constant $\alpha$ is treated in the Thompson limit, corresponding to the use of the coupling evaluated at zero momentum transfer $\alpha(0)$. In contrast, ref.~\cite{Jiang:2024dhf} employed the $G_\mu$ scheme. Both computations report QED corrections of the order of $1\%$, largely independent of the species of the colliding hadrons or their energies. Surprisingly, however, the weak corrections in the $G_\mu$ scheme are found to contribute an additional $-4$ to $-5\%$ to the total cross section, an unexpected result given that the hard scale of the process, $W_{\gaga}^2\sim M_\tau^2 \ll M_W^2$, suggests weak corrections should be suppressed. A more recent calculation in a hybrid scheme confirmed that weak corrections are indeed negligible and that the full NLO EW corrections are close to the pure QED ones~\cite{Shao:2025bma}. Thereby, in the hybrid scheme, couplings to quasi-real photons (such as the coherent initial-state photons) are renormalised in the $\alpha(0)$ scheme, while other couplings (e.g. those to photons from real-emission) are renormalised in the $G_\mu$ scheme, making it closely aligned with the $\alpha(0)$ scheme.

The scheme dependence was further investigated in ref.~\cite{Dittmaier:2025ikh}, where both $\alpha(0)$ and $G_\mu$ schemes were applied. Once again, large weak corrections were observed in the $G_\mu$ scheme, whereas the $\alpha(0)$ scheme gave only the expected $\mathcal{O}(1\%)$ effects. This discrepancy arises from the fact that in the $G_\mu$ scheme the LO cross section is overestimated, leading to large spurious NLO corrections in order to reproduce the physical result. These studies clearly indicate that the hybrid approach should be preferred for UPC predictions to ensure stable perturbative behaviour.

\begin{figure}
    \includegraphics[width=0.49\textwidth]{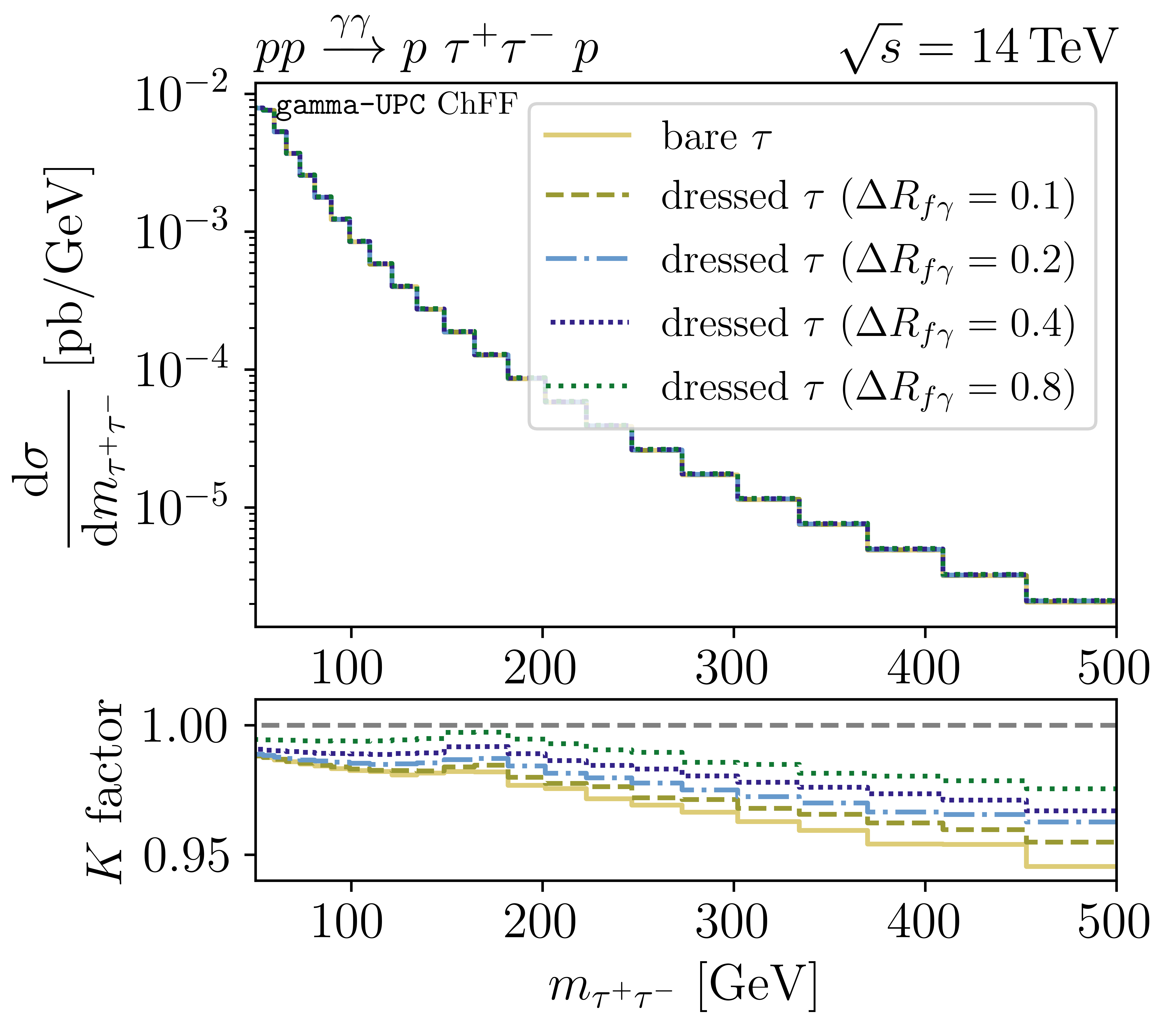}
    \includegraphics[width=0.49\textwidth]{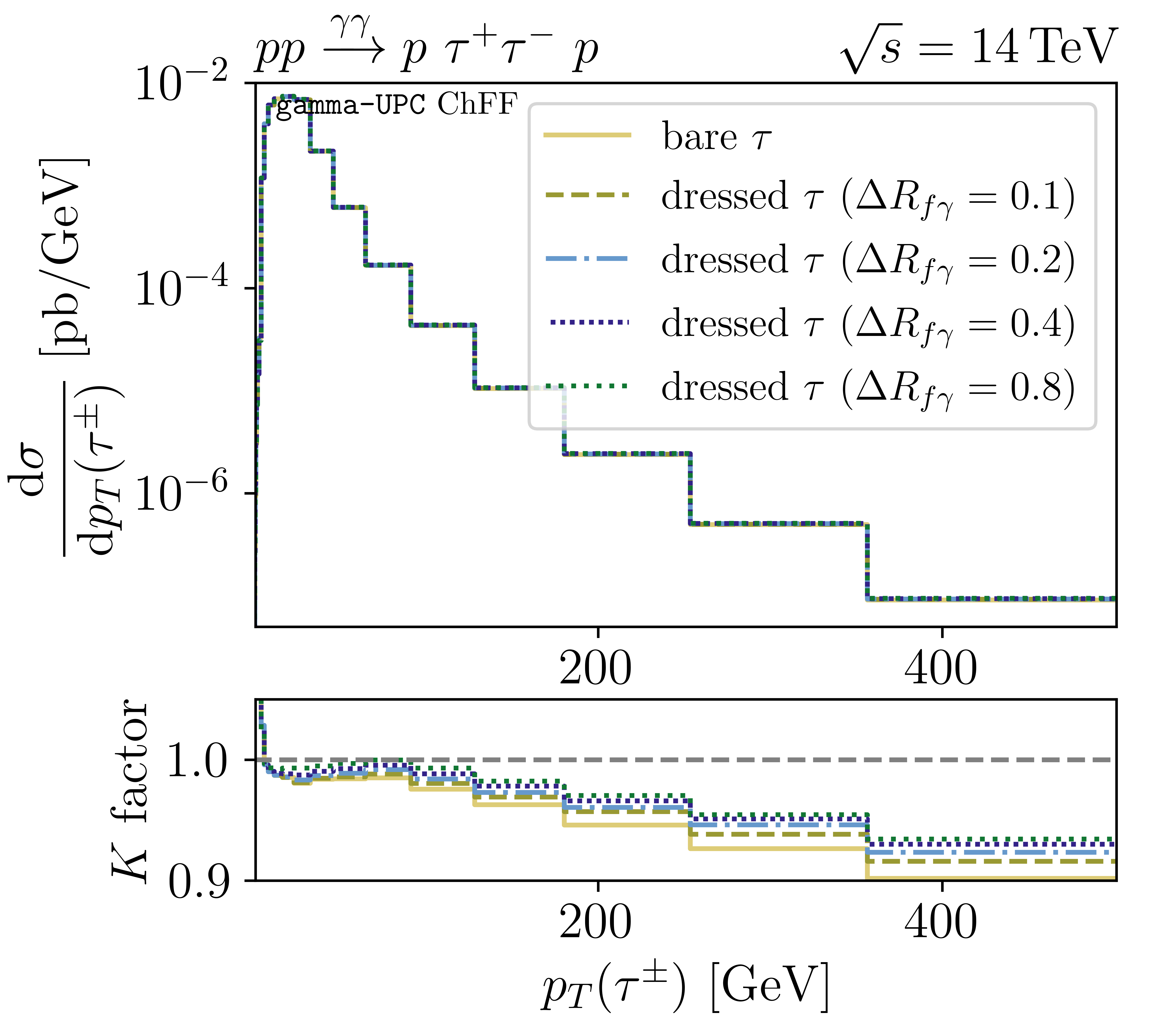}
    \caption{Differential cross sections for $\tau$-pair production as functions of the invariant mass, $\mathrm{d}\sigma/\mathrm{d}m_{\tata}$~(left), and the transverse momentum of $\tau^\pm$, $\mathrm{d}\sigma/\mathrm{d}p_{T}(\tau^\pm)$ (right), in p-p UPCs at $\sqrt{s}=14\,\mathrm{TeV}$, with the fiducial cuts $m_{\tata}>50\,\mathrm{GeV}$ and $|\eta(\tau^\pm)|<2.5$. The lower panels show the $K$ factors. The distributions for the bare $\tau$ are compared to those for the dressed $\tau$ lepton, defined with four different values of $\Delta R_{f\gamma}$. Figure taken from ref.~\protect\cite{Shao:2025bma}.}
    \label{fig:aa2tautau}
\end{figure}

The framework developed in ref.~\cite{Shao:2025bma}, which combines the \gammaUPC\ code with \mgshort\ and enables automated NLO calculations for UPC processes, makes computations like those presented in this section straightforward. As an illustrative example, it was used to compare the behaviour of bare and dressed leptons. A dressed lepton is defined through a photon recombination, where any photon within an angular distance $\Delta R_{f\gamma}$ of a lepton is merged with the nearest lepton. This procedure allows the lepton mass to be neglected without introducing infrared divergences. The invariant-mass distribution of the $\tau$ pair (with additional fiducial cuts $m_{\tata} > 50\,\mathrm{GeV}$ and $|\eta(\tau^\pm)| < 2.5$, analogous to those used by CMS~\cite{CMS:2024qjo}), shown in \cref{fig:aa2tautau}, demonstrates that dressing the $\tau$ significantly reduces the NLO corrections in the high invariant-mass region. This highlights how quasi-collinear logarithmic enhancements can be mitigated by lepton dressing and shows that increasing the isolation radius leads to a smoother and more uniform $K$-factor across the distribution.

\subsection{$\gaga \to \gaga$}\label{subsec:gaga}

Another interesting process we discuss here is LbL scattering, $\gaga \to \gaga$. Computing NLO corrections to this process is challenging, as the LO contribution is loop-induced, implying that NLO computations involve two-loop Feynman diagrams. Approximations for these corrections in the high-~\cite{Bern:2001dg,Binoth:2002xg} and low-energy limits~\cite{Martin:2003gb} were obtained more than two decades ago. However, the need for exact NLO results has become pressing since ATLAS reported a $2\sigma$ deviation in Pb-Pb UPCs at $\sqrt{s_{NN}}=5.02\,\mathrm{TeV}$ between the LO predictions and fiducial cross-section measurements~\cite{ATLAS:2020hii}.

\begin{figure}[t!]
\includegraphics[height=8.1cm,valign=c]{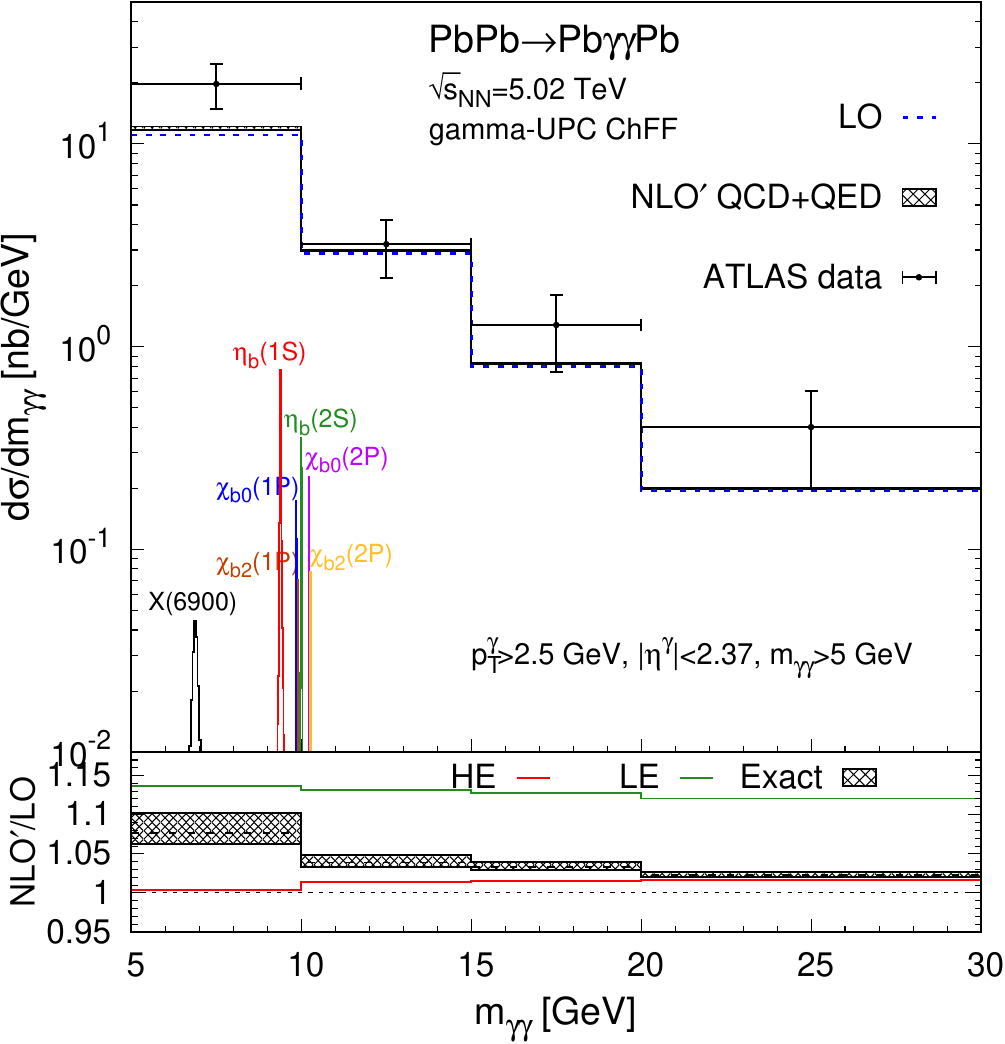}
\includegraphics[height=9.2cm,valign=c]{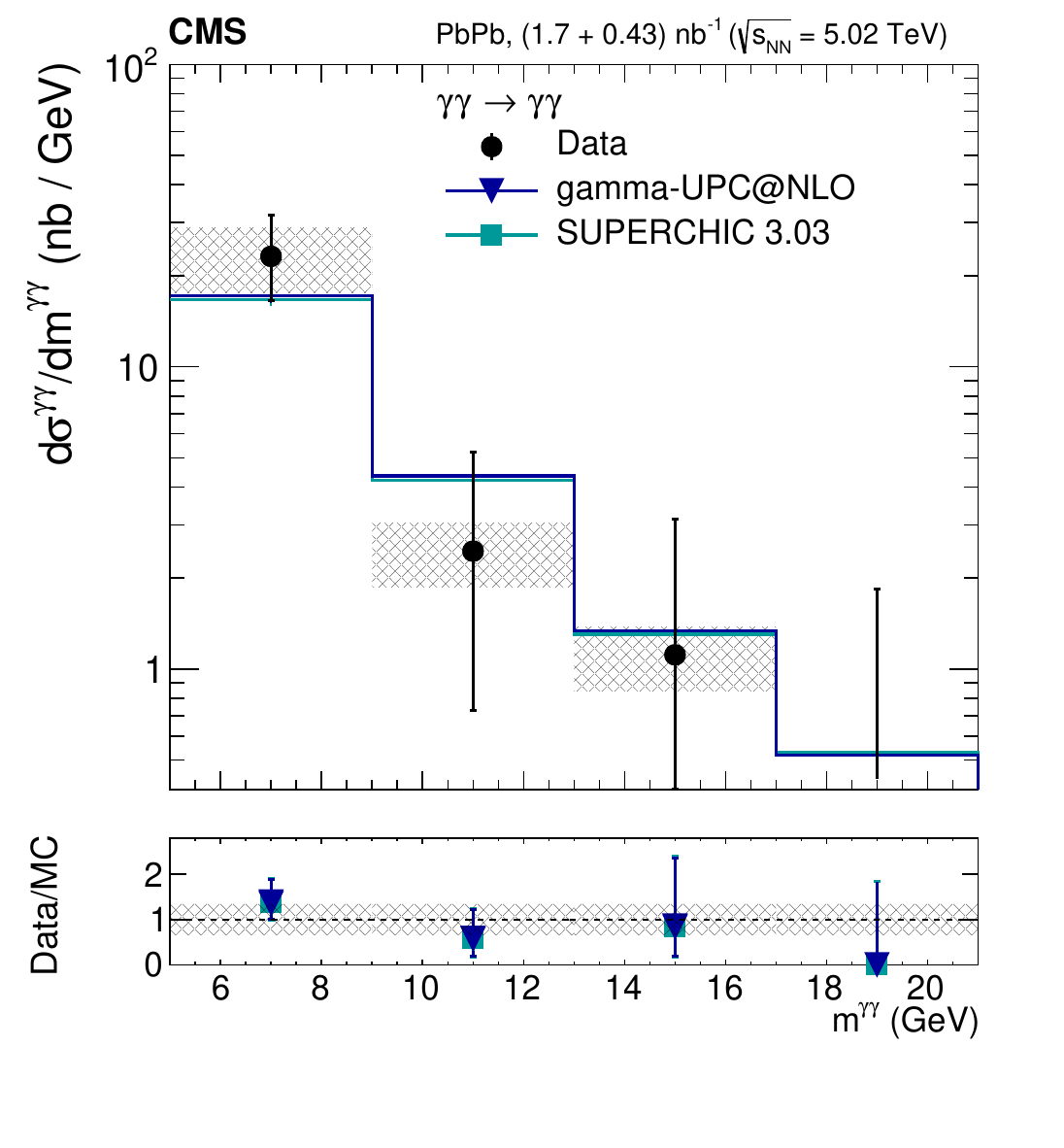}
\caption{\textit{Left:} Comparison between the LO (blue dashed), NLO$^\prime$ QCD+QED (grey hatched) with the ATLAS measurement~\cite{ATLAS:2020hii} for the photon pair's invariant-mass distribution. \textit{Right:} Comparison between NLO$^\prime$ QCD+QED predictions with \gammaUPC\ (blue triangles) and LO \superchic\ predictions (grey boxes) with the CMS measurement (black dots)~\protect\cite{CMS:2024bnt}. Figures taken from ref.~\protect\cite{AH:2023kor} (left) and ref.~\protect\cite{CMS:2024bnt} (right).}
\label{fig:aa2aa}
\end{figure}

Exact NLO QCD and QED corrections with full fermion-mass dependence have been computed using two independent strategies. The first relies on an analytical approach utilising differential equations and a canonical basis~\cite{AH:2023ewe}. The second is fully numerical and employs the Local Unitarity (LU) method~\cite{Capatti:2020xjc,Capatti:2022tit}. The agreement between the two results provides a strong cross-check for both calculations. Moreover, these predictions have been directly compared to the ATLAS measurement~\cite{ATLAS:2020hii} in ref.~\cite{AH:2023kor}. The photon pair's invariant-mass spectrum is shown in the left panel of \cref{fig:aa2aa}. Using the so-called NLO$^\prime$ prediction -- where the prime indicates that partial NNLO effects are included to enhance accuracy -- it was found that the inclusion of exact NLO QCD and QED corrections reduces, but does not fully resolve, the moderate tension observed in LbL scattering. In contrast, the CMS measured invariant mass spectrum in Pb-Pb UPCs at $\sqrt{s_{NN}}=5.02\,\mathrm{TeV}$ matches the theory predictions from both \superchic\ at LO and ref.~\cite{AH:2023kor} at NLO$^\prime$ well within the quoted uncertainties, see the right panel of \cref{fig:aa2aa}~\cite{CMS:2024bnt}. 

%%%%%%%%%%%%%%%%%%%%%%%%
%% SECTION 3          %%
%%%%%%%%%%%%%%%%%%%%%%%%
\section{Automation of NLO Computations}

Some results in the previous sections were obtained with the automated framework for NLO computations for any UPC two-photon process, based on \gammaUPC\ and \mgshort. The main advantage of this tool is its accessibility, it enables NLO predictions to be produced quickly and reliably, even by non-experts. In the following we briefly introduce the framework, outline its features, and point out some of its current limitations. A more detailed description can be found in ref.~\cite{Shao:2025bma}.

The tool allows one to compute UPC processes in arbitrary proton and nucleus collisions in a straightforward manner. Fixed-order calculations include all LO and NLO terms as introduced in \cref{{eq:decomposition_lo},{eq:decomposition_nlo}}. These computations can be carried out with only a few input commands, closely following the syntax of the standard \mgshort\ implementation~\cite{Alwall:2014hca,Frederix:2018nkq}. Incoming photons are specified by the symbol \texttt{!a!}, where the exclamation marks indicate coherent photons and distinguish them from the partonic photon, denoted by \texttt{a}. As an example, the syntax for generating a process with four (multi)particles {\tt p$_i$} in the spectrum of the NLO Universal Feynman Output (UFO)~\cite{Degrande:2011ua,Darme:2023jdn} model \texttt{myNLOmodel\_w\_qcd\_qed} is
\vskip 0.25truecm
\noindent
~~\prompt\ {\tt ~import~model~myNLOmodel\_w\_qcd\_qed-restrict\_card\_w\_a0}

\noindent
~~\prompt\ {\tt ~generate~!a! !a! > p$_1$ p$_2$ p$_3$ p$_4$ [QCD QED]}

\vskip 0.25truecm
\noindent
where the first line imports the model with the \texttt{a0} or \texttt{Gmu} restriction. The restriction is necessary to automatically adopt the hybrid $\alpha$ renormalisation scheme, where couplings to quasi-real photons are treated in the $\alpha(0)$ scheme and other couplings in a scheme that allows for a proper UV renormalisation at NLO such as the $G_\mu$ scheme or the $\alpha(M_Z)$ scheme.\footnote{Hybrid renormalisation UFO models were first introduced into \mgshort\ in the context of final states with tagged photons~\cite{Pagani:2021iwa}. Further details can be found there and in ref.~\cite{Shao:2025bma}.} Once a process is generated, the computation proceeds with the usual \texttt{output} and \texttt{launch} commands. The corresponding settings in the \texttt{run\_card.dat} file must then be adjusted for UPC input. Beam types should be set to \texttt{lpp1=2} and \texttt{lpp2=2}. The \gammaUPC\ photon flux can be chosen as EDFF or ChFF via \texttt{pdlabel=edff} or \texttt{pdlabel=chff}, respectively.\footnote{For p–p collisions, the option \texttt{pdlabel=iww} uses the improved Weizsäcker–Williams approximation~\cite{Frixione:1993yw}.} By default, \mgshort\ assumes p-p collisions, for heavy-ion collisions, the number of protons and neutrons in each beam can be explicitly specified using the key words \texttt{nb\_proton1} and \texttt{nb\_neutron1} for the hadron of the first beam. For the second beam the key words are simply \texttt{nb\_proton2} and \texttt{nb\_neutron2}. It is important to note that in heavy-ion collisions, the beam energy corresponds to the total ion energy and not the energy per nucleon. An example run card for a p–Pb UPC at $\sqrt{s_{NN}}=8.8\,\mathrm{TeV}$ is shown in \cref{lst:run_card}.

\begin{lstlisting}[caption={Excerpt of the \texttt{run\_card.dat} settings for a p–Pb UPC at $\sqrt{s_{NN}}=8.8\,\mathrm{TeV}$.},captionpos=b,language=Python,label={lst:run_card},float=tp,floatplacement=tbp]
#***********************************************************************
# Collider type and energy                                             *
#    0 = no PDF                                                        *
#    1/-1 = proton/antiproton                                          *
#    2 = elastic photon of proton/ion beam                             *
#    3/-3 = electron/positron with ISR/Beamstrahlung;                  * 
#    4/-4 = muon/antimuon with ISR/Beamstrahlung;                      * 
#***********************************************************************
  2     = lpp1 # beam 1 type (0 = no PDF)
  2     = lpp2 # beam 2 type (0 = no PDF)
  7000.0    = ebeam1 # beam 1 energy in GeV
  574080.0  = ebeam2 # beam 2 energy in GeV
#*********************************************************************
# Note that ebeam1 and ebeam2 are energies of the ion beams          *
# instead of energies per nucleon in nuclei                          *
# For instance, the LHC beam energy of 2510 GeV/nucleon in Pb208     *
# should set 2510*208=522080 GeV for ebeam                           *
#*********************************************************************
  1     = nb_proton1  # number of proton for the first beam
  0     = nb_neutron1 # number of neutron for the first beam
  82    = nb_proton2  # number of proton for the second beam
  126   = nb_neutron2 # number of neutron for the second beam
#***********************************************************************
# PDF choice: this automatically fixes also alpha_s(MZ) and its evol.  *
# pdlabel: lhapdf = LHAPDF (installation needed) [1412.7420]           *
#          iww = Improved Weizsaecker-Williams Approx.[hep-ph/9310350] *
#          edff = EDFF in gamma-UPC            [eq.(11) in 2207.03012] *
#          chff = ChFF in gamma-UPC            [eq.(13) in 2207.03012] *
#***********************************************************************
  chff  = pdlabel # PDF set
\end{lstlisting}

The NLO UPC implementation integrates seamlessly into the \mgshort\ framework and can therefore be combined with other features of the code. A particular useful example is the interface to standard Monte Carlo tools such as \herwig~\cite{Corcella:2000bw,Corcella:2002jc,Bahr:2008pv,Bellm:2013hwb,Bellm:2015jjp} or \pythia~\cite{Sjostrand:2006za,Bierlich:2022pfr} via the Les Houches Event output format~\cite{Alwall:2006yp}, which allows further event processing. Consequently, not only fixed-order computations but also parton shower (PS) simulations become possible. While LO+PS computations can be performed without restrictions, at NLO+PS only processes with NLO QCD can currently be linked to the showering tools via the MC@NLO method~\cite{Frixione:2002ik}. Another limitation is that NLO+PS calculations are restricted to \textit{processes without jets} at Born level. Of course, jets may still appear through real-emission contributions at NLO. A summary of the available options is given in \cref{tab:allowedinMG}, where fLO and fNLO denote fixed-order LO and NLO computations, respectively.

\begin{table}[t!]
\centering
\tabcolsep=3.5mm
\vspace{0.2cm}
\begin{tabular}{|l|c|c|c|c|} \hline
 & \multicolumn{4}{c|}{\gammaUPC+\mgshort}\\
 & \multicolumn{2}{c|}{Processes without jets} & \multicolumn{2}{c|}{Processes with jets}\\
 & QCD ($i=1$) & EW ($i\geq 2$) & QCD ($i=1$) & EW ($i\geq 2$) \\
\hline
fLO &  \cmark &  \cmark & \cmark & \cmark \\\hline
LO+PS &  \cmark &  \cmark & \cmark & \cmark \\\hline
fNLO &  \cmark &  \cmark & \cmark & \cmark \\\hline
NLO+PS &  \cmark &  \xmark & \xmark & \xmark \\
\hline
\end{tabular}
\caption{Overview of the currently available (\cmark) and unavailable (\xmark) types of computations within the \gammaUPC+\mgshort\ framework for photon–photon processes in UPCs. The classification into \textit{with} or \textit{without jets} refers to the Born-level configuration; jets may still be produced at NLO through real radiation even in processes labeled \textit{without jets}. The index $i$ corresponds to the $\Sigma_{\mathrm{(N)LO}_i}$ terms in \cref{eq:decomposition_lo,eq:decomposition_nlo}.
\label{tab:allowedinMG}}
\end{table}

\begin{figure}[t!]
   \includegraphics[width=0.49\textwidth]{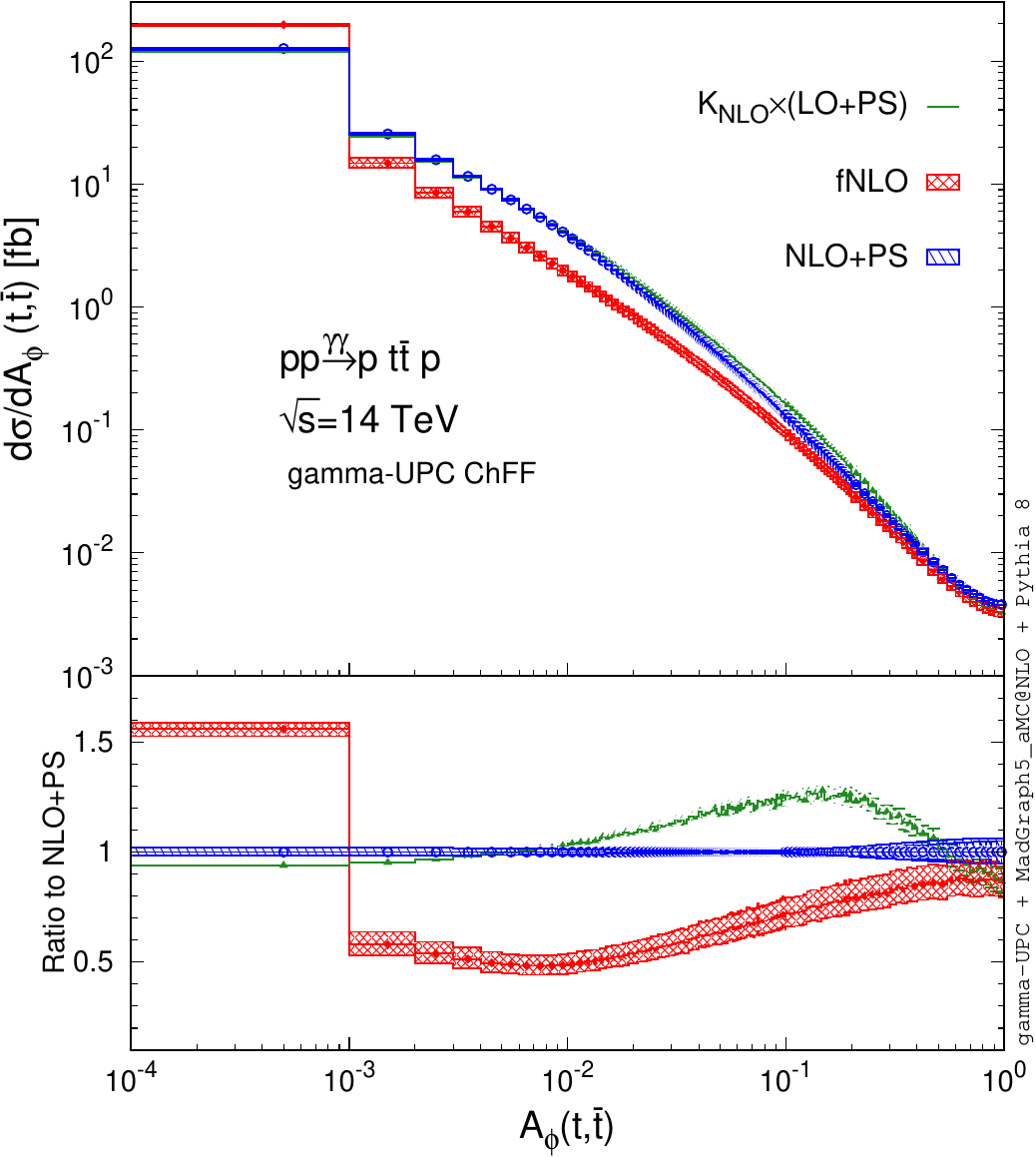}
   \includegraphics[width=0.49\textwidth]{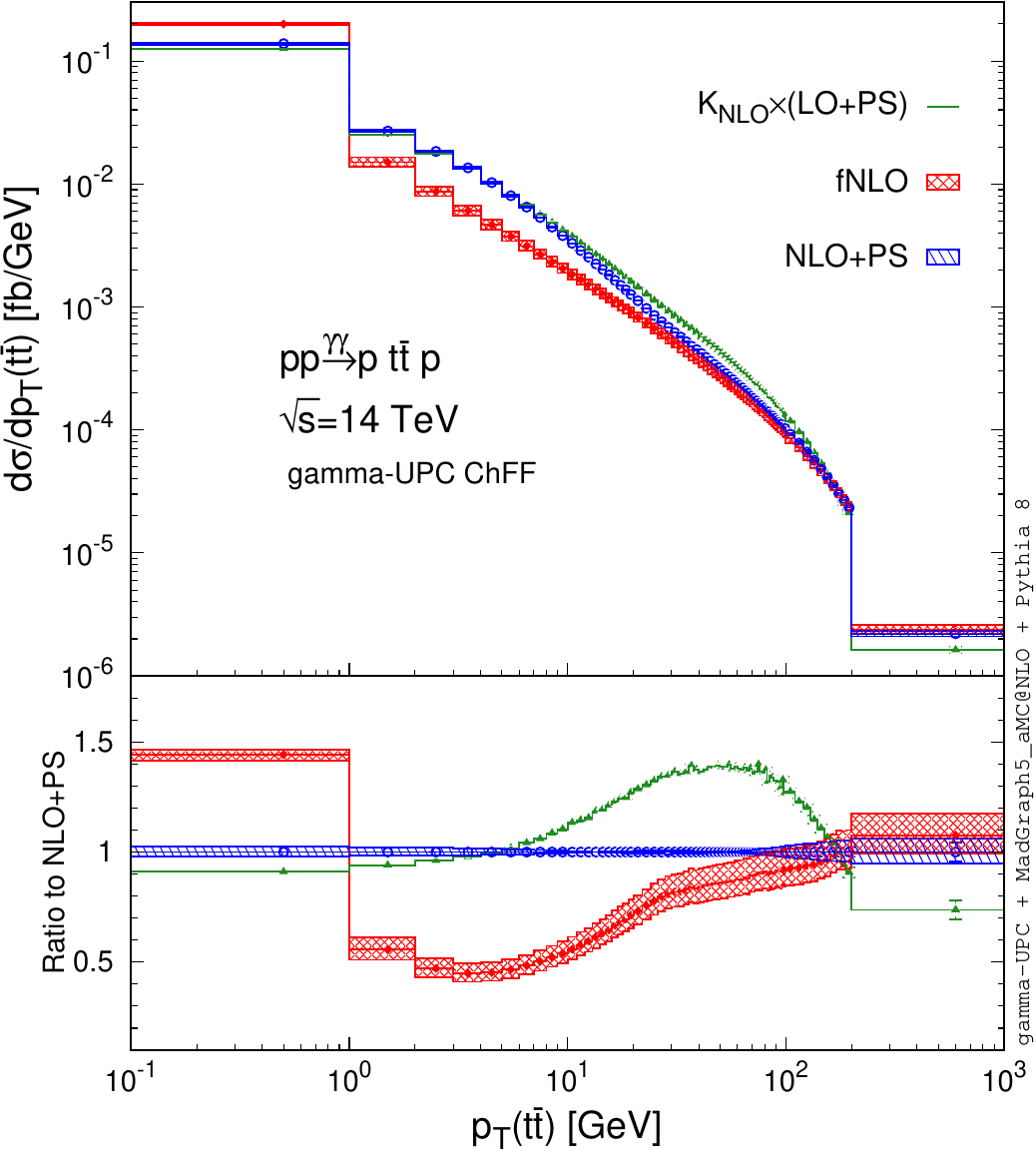}
   \caption{Differential distributions of the top-quark pair acoplanarity, $\mathrm{d}\sigma/\mathrm{d}A_{\phi}(t,\bar{t})$ (left), and transverse momentum, $\mathrm{d}\sigma/\mathrm{d}p_T(\ttbar)$ (right), in p–p UPCs at $\sqrt{s}=14,\mathrm{TeV}$, computed using the ChFF $\gamma$ fluxes implemented in \gammaUPC. Results are shown at LO+PS (green histograms), fixed-order NLO (fNLO, red histograms), and NLO+PS (blue histograms) accuracy, where NLO refers to NLO QCD corrections only. The LO+PS distributions are rescaled by the overall $K$ factor, $K_{\mathrm{NLO}}\equiv\sigma_{\mathrm{NLO}\ \mathrm{QCD}}/\sigma_{\mathrm{LO}}\approx 1.21$. The lower panels display ratios relative to the NLO+PS central values. Hatched bands represent renormalization-scale uncertainties for fNLO and NLO+PS, while error bars indicate statistical uncertainties. The first bins in both spectra extend to zero. Figure taken from ref.~\protect\cite{Shao:2025bma}.}
    \label{fig:aa2ttbarNLOPS}
\end{figure}

Despite these limitations, the framework already covers a wide class of phenomenologically relevant processes. As an example, \cref{fig:aa2ttbarNLOPS} shows the NLO+PS prediction for the process $\gaga \to \ttbar$ in p-p UPCs at $\sqrt{s}=14\,\mathrm{TeV}$. Here, only the NLO QCD corrections are included, which are the dominant higher-order contributions, enhancing the cross section by about $\mathcal{O}(+20\%)$ compared to LO. The figure shows the differential cross section as a function of the top-quark pair acoplanarity, $A_\phi(t,\bar{t})=1-|\Delta\phi(t,\bar{t})|/\pi$, where $\Delta\phi(t,\bar{t})$ denotes the azimuthal separation between the top and anti-top, and of the pair’s transverse momentum, $p_T(\ttbar)$. In both observables, the PS has a pronounced impact in the low-$A_\phi(t,\bar{t})$ and low-$p_T(\ttbar)$ regions, where it modifies the fNLO prediction by up to $\mathcal{O}(50\%)$. Such effects are therefore essential when aiming to match theory predictions with experimental data.

%%%%%%%%%%%%%%%%%%%%%%%%
%% CONCLUSIONS        %%
%%%%%%%%%%%%%%%%%%%%%%%%
\section{Conclusions}

The study of UPCs is a vibrant and rapidly evolving field. They provide exceptionally clean and unique experimental signatures, allowing the investigation of phenomena that remain hidden in inclusive high-energy collisions. On the theoretical side, predictions for photon-photon UPCs at hadron colliders have advanced significantly. The modelling of increasingly realistic photon fluxes continues to improve, and the transition from LO to NLO accuracy within collinear factorisation has recently been achieved, establishing NLO computations as the new standard. Many NLO corrections have already been worked out in a process-specific or tailored manner. A particular remarkable result is the computation of NLO QCD+QED corrections to $\gaga \to \gaga$ with exact fermion-mass dependence, involving two-loop Feynman diagrams with multiple scales. This calculation is the first entirely new result obtained with the novel multi-loop approach based on the Local Unitarity (LU) construction, thereby demonstrating the potential of this fully numerical approach in a broader context.

A major step forward for exclusive photon-photon UPCs is the introduction of the automated \gammaUPC+\mgshort\ framework. It allows even non-expert users to carry out studies for $\gaga$ collisions at hadron colliders with NLO precision including QCD as well as EW corrections. With only a few lines of input, it is possible to generate processes, simulate events, and interface them to parton shower (PS) tools such as \herwig\ or \pythia. Current limitations remain, for instance NLO+PS simulations are available only for pure QCD corrections and for process without jets at Born level. Furthermore, the automated framework presently supports only elementary particles, excluding bound states such as quarkonia. Nevertheless, the theoretical accuracy for processes in photon-photon collisions starts to catch up with that for inclusive parton-parton scattering processes.

\section*{Acknowledgments}

This work is supported by the ERC (grant 101041109 ``BOSON") and the French ANR (grant ANR-20-CE31-0015, ``PrecisOnium"). Views and opinions expressed are however those of the authors only and do not necessarily reflect those of the European Union or the European Research Council Executive Agency. Neither the European Union nor the granting authority can be held responsible for them.

\bibliography{references}

\end{document}